# Enhancing Path-Oriented Test Data Generation Using Adaptive Random Testing Techniques

Esmaeel Nikravan, Farid Feyzi, Saeed Parsa
Faculty of computer engineering
Iran University of Science and Technology
Tehran, Iran
nikravan@comp.iust.ac.ir, farid_feyzi@comp.iust.ac.ir, parsa@iust.ac.ir

*Abstract*—in this paper, we have developed an approach to generate test data for path coverage based testing. The main challenge of this kind testing lies in its ability to build efficiently such a test suite in order to minimize the number of rejects. We address this problem with a novel divide-and-conquer approach based on adaptive random testing strategy. Our approach takes as input the constraints of an executable path and computes a tight over-approximation of their associated sub-domain by using a dynamic domain partitioning approach. We implemented this approach and got experimental results that show the practical benefits compared to existing approaches. Our method generates less invalid inputs and is capable of obtaining the sub-domain of many complex constraints.

*Abstract—software testing; test data; path coverage; random testing; adaptive random testing.*

## I. INTRODUCTION

Software testing is one of the most important and practical techniques to ensure software quality. One dilemma task of software testing is to select test cases that effectively detect faults at a minimum cost. Many testing approaches have been developed to guide the test data generation [1-6]. One simple and common method is *Random Testing* (RT) [7-9], in which test data are selected in a random manner from the program's input domain. The main criticize that exist against random testing is that does not uses of any information from the program under the test Similar to RT, *adaptive random testing* (ART) [10-15] also randomly generates program inputs from the input domain this strategy has been proposed to improve the performance of RT in terms of using fewer test cases to detect the first failure. ART makes use of additional criteria to choose inputs as test cases in order to evenly spread test cases over the input domain. One of well-known software testing techniques is *Path testing*. The basic idea in path testing is to find at least one test data to activate each selected path. Each path is associated with a *path condition* that is the conjunction of all the predicate interpretations that are taken along the path.

The path condition represents the constraints that have to be satisfied for inputs in order to execute the path. Finding an exact solution set to complete solving a path condition is NP-hard [21]. Hence, the test data generation, is considered a major challenge in path testing. So, in this paper, we present a path-oriented automatic random testing method based on ART. The approach gets the constraint set of the input variables for exercising a path, and then a dynamic domain partitioning approach is used to compute the sub-domain of input variables along a chosen path. The experiment results show that the domain gotten by our method is more accurate than the PRT, and random testing efficiency can thus be enhanced by using the proposed method. The remainder of this paper is organized as follows: Section two gives a motivating example. Section three describes random testing and path-oriented random testing strategy. Section four present our random test data generation method based on adaptive random testing. Section five reports experimental results to show that the method is effective and practicable; finally, the conclusions are presented in Section six.

## II. Motivating Example

Consider the C program foo showed in Fig.1 and the problem of generating test data for path 1→2→3→4→5→6.

|   | int foo(int x, int y) {         |
|---|---------------------------------|
| 1 | if (y<= 8*sin(0.2*x+7)+4) {     |
| 2 | …                               |
| 3 | if (y<= sqrt(x)+8){             |
| 4 | ...                             |
| 5 | if (x<=16-y){                   |
| 6 | ...                             |
|   | }}}                             |

Fig1. Example Program foo

Let us use a random test data generation method to generate pairs $(x_i, y_i)$ in $0..15 \times 0..15$. Inputs which satisfy the below path condition related to mentioned path will be accepted and other pairs will be rejected.

$$y \leq 8 * sin(0.2 * x + 7) + 4 \quad (1)$$
$$and (y \leq sqrt(x) + 8) and (x \leq 16 - y)$$

However, using RT is highly expensive as it generates a large number of invalid inputs and will reject them. In fact, by manually analyzing the program, we can see that the average probability of rejecting a pair is about %52 with this approach.

Symbolic analysis could be used to reduce the number of invalid inputs. To this aim, a *constraint solver* is required to solve the path condition and generate satisfying inputs. However, checking constraint satisfiability is undecidable and symbolic analysis potency is limited by the power of constraint solver. In fact, constraint solvers are unable to handle non-linear and very complex constraints, such as (1).

### III. RELATED WORK

Random testing is a basic and simple software testing technique, which selects test cases at random from the set of all possible program inputs. This method selects test cases, according to a uniform distribution strategy, that is, all program inputs have the same probability to be chosen as test cases. For example, for a program with two input variables *x* and *y*, their input domain D can be represented as $D=D_x \cup D_y$, where $D_x$ ($D_y$), called variable domain, is a set of all values that input variable *x(y)* can hold. RT can be implemented just by selecting *x* and *y* from its domain at random, respectively, meaning when selecting *y* without paying attention on the value obtained for *x*. In other word, the two variable values are independently determined. Obviously, if the domain of *x* or that of *y* can be reduced, the test generation on the invalid domain can be avoided. Therefore, a key question of RT is how to get a precise input domain. If it is difficult to obtain a precise input domain, we hope to get the most approximate solution to the one.

*Path Random Testing* (PRT) that is proposed by Gotlieb et. al.[18-20] works like random testing with this difference that it selects test data at random to cover a given subset of paths according to a uniform probability distribution over the program's input domain. More specifically, PRT applied constraint propagation to get input domain along a given path, and then a path-oriented random test data generation was performed. The goal of constraint propagation was to shrink the finite variation domain of each variable in order to get an approximation of the solutions with respect to a set of constraints. The PRT algorithm took as inputs a set of variables, a constraint set corresponding to the path conditions of the selected path, and a division parameter *k* (a given parameter). The algorithm separated each variable domain into *k* equal sub-domains. If the size of a variable domain could not be divided by *k*, the domain was enlarged until its size could be divided by *k*. By iterating this process over all the *n* input variables, the input domain would be partitioned into $k^n$ sub-domains. The sub-domains that could not satisfy the path constraints would be omitted. As a result, some invalid inputs were removed, so the test generation efficiency could be increased.

Chan et al. [10-15] pointed out that failure-causing inputs tend to be clustered within the input domain. They even roughly classified failure patterns (i.e. patterns of failure causing inputs within the input domain) into the categories block, strip, and point. Based on this observation, they introduced Adaptive Random Testing (ART) which is designed to *evenly spread* test cases, because two nearby test cases have a high probability of either detecting no failure or detecting the same failure (pattern). As the concept of even spread can be implemented in different ways, several ART methods (algorithms) have been proposed [10-15]. Each of these approaches has its own strengths and weaknesses, regarding runtime and testing effectiveness.

Distance-based ART (D-ART) and Restriction-based ART (R-ART) are the first two attempts, which have significantly improved the fault-detection capability of RT. To further reduce the overhead of ART while keeping a high fault-detection capability, Chen et al., introduce a new ART method, namely ART through Iterative Partitioning (IP-ART) [16-17]. Conventionally, partitioning is a strategy to group elements having similar behaviors in some sense into the same sub-domain. IP-ART uses partitioning to identify a test case generation region, where inputs have higher chance of being far apart from all successful test cases. If such a test case generation region cannot be identified under current partitioning scheme, the input domain will be repartitioned using a finer partitioning scheme. Since IP-ART does not require the generation of extra candidates and avoids the distance computations and comparisons, it has much lower computational overhead while keeping a high fault-detection capability comparable to that of D-ART and R-ART. Based on the similarity that exists between failure patterns and path domain, we propose an approach (inspires from IP-ART and PRT) that obtain a tight over-approximation of sub-domain that values of it satisfy the path constraints.

### IV. THE PROPOSED APPROACH

This section presents the main idea of our proposed approach. We present a method that performs PRT based on ART. The method takes as inputs a set of variables along with their variation domain, PC a constraint set corresponding to the path conditions of the selected path. We need to first decide the resolution of the grid for partitioning the input domain. The granularity of the grid has a severe effect on speed of search and precision of our method and obtained sub-domains. This is because, if the grid is too fine at the beginning, then many sub-domains will not be sufficiently far away from the invalid sub-domains (sub-domains that don't satisfy the PC). Hence, the algorithm starts with a coarse grid. If no valid domain is found and no candidate sub-domain (sub-domain that neither invalid nor adjacent) is available, then the current $n \times n$ partitioning scheme will be discarded and a finer partitioning scheme using an $(n+1) \times (n+1)$ grid will be applied to partition the input domain all over again. On the other hand coarse grained grid may give a sub-domain that it has till many invalid values that don't satisfy the PC.

To illustrate the basic idea of our approach, let us consider Fig. 2. It shows a square input domain of variables x, y such that $x, y \in [0..15]$. Assuming that the input domain is partitioned using a 4×4 grid. As a result, we get the 16 following sub-domains:

$D_1 = (x \in [0..3], y \in [12..15])$,

$D_2 = (x \in [0..3], y \in [8..11])$,

…

$D_{16} = (x \in [12..15], y \in [0..3])$

That form a partition of the (augmented) hypercuboid D = $(x \in 0..15, y \in 0..15)$

We know that the valid sub-domains that satisfy PC are $D_2, D_3, D_4, D_6, D_7, D_8, D_{11}, D_{12}, D_{16}$.

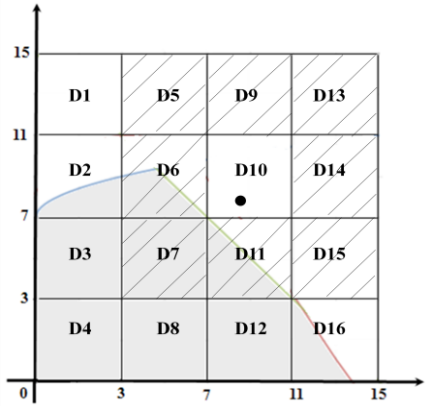

Fig. 2 Partitioning the input domain

A. Finding first valid domain

After partitioning the input domain space into smaller sub-domains, one of them, say, $D_{10}$, is randomly selected as the region for generating the first test case. A test case is randomly generated within this region, and sent to a function that checks satisfaction of PC with this test case. If the PC is satisfied with this test case, then we found the first valid sub-domain, and we go to the next step. In our example the test data that generated from sub-domain $D_{10}$ can not satisfy the PC. Then the adjacent sub-domains surrounding $D_{10}$ are marked as shown. So the sub-domains $D_1, D_2, D_3, D_4, D_8, D_{12}, D_{16}$ are now the only remaining candidate cells and, therefore, the second test case will be generated from this region. This process will be repeated until a valid sub-domain is detected or no sub-domains are left for consideration. If no accepted sub-domain is found and no candidate sub-domain is available, then the current 4×4 partitioning scheme will be discarded and a finer partitioning scheme using an 5×5 grid will be applied to partition the input domain all over again.

B. Path constraint testing function

This function takes as input generated values of variables, as test cases, involved in path constraint. It consists of an *if-then* statement. The condition part of *'if'* statement is the PC and it returns TRUE if the constraints are satisfied, otherwise returns FALSE. Fig.3 represents the path constraint testing function for motivating example.

---
BOOL TEST(int x,y)
{
Return((y<= 8*sin(0.2*x+7)+4) &&(y<= sqrt(x)+8) && (x<=16-y))
}
---

Fig.3 path constraint testing function

C. Finding other valid sub-domains

After finding the first valid sub-domain, we can examine its neighbours for more valid sub-domains. To this end, we choose a test case from each of neighbor sub-domains and pass it to *path constraint testing function* to check whether it satisfies the path condition or not. It is important to note that the test case must be chosen from a region of a neighbor sub-domain that has common boundry with pre-identified valid sub-domain. When the function returns True, the sub-domain is considered as valid, otherwise it will be discarded.

V. EMPIRICAL RESULTS

We implemented proposed approach and compared it with RT and PRT [18]. All implementations take path conditions and domains as input parameters and provide a uniform random test suite as a result. To be fair, all implementations (PRT, RT and proposed approach) make use of the same random number generator. Our approach and PRT also come with an additional parameter n which is the grid resolution parameter defined in Section IV. When n = 1, the input domain is not divided and both proposed approach and PRT will performed same as RT. When n > 1, the domain space is partitioned into several sub-domains according to *n* and the method seeks to the first valid sub-domain using IP-ART method.

We evaluated our approach w.r.t PRT and RT on the foo program given in Fig. 1. All the experimental results were computed on a 2.4 GHz Intel Core Duo with 4GB of RAM. Fig. 4 reports on the results obtained for the path $1 \rightarrow 2 \rightarrow 3 \rightarrow 4 \rightarrow 5 \rightarrow 6$ in the foo program by regularly increasing the desired length of the random test suite. Fig. 4 shows the number of test data generated with the our approach with three distinct values of the grid resolution and other two approachs, PRT and traditional RT. For example, the first column shows that the number of rejects of the RT method is $240 − 100 = 140$ test data and $123-100=23$ test data for PRT when n=4, 20 when n=5, and so on; While it evaluates to $123 − 100 = 23$ with proposed approach when n = 4, 19 when n = 5, and so on.

| Requested | 100 | 500 | 1000 | 2000 | 5000 | 10000 |
|---|---|---|---|---|---|---|
| RT | 240 | 1108 | 2127 | 4279 | 10325 | 21029 |
| PRT(n=4) | 123 | 602 | 1209 | 2413 | 6025 | 11978 |
| PRT(n=5) | 120 | 591 | 1185 | 2372 | 5995 | 11901 |
| PRT(n=6) | 118 | 579 | 1160 | 2295 | 5752 | 11604 |
| Proposed(n=4) | 123 | 591 | 1201 | 2392 | 6994 | 11607 |
| Proposed(n=5) | 119 | 583 | 1160 | 2327 | 5901 | 11516 |
| Proposed(n=6) | 115 | 571 | 1154 | 2249 | 5704 | 11402 |

Fig. 4. Length of the test suite generated for foo.

To compare the capability of proposed approach with that of existing methods, a number of experiments have been conducted. We compare proposed approach with *Pex*. Pex [22] is a tool that implements Dynamic Symbolic Execution to generate test inputs for .NET code, supporting languages such as C#, Visual Basic, and F#. our experiment shows that for many of non-linear constraints that PEX cannot solves them, proposed approach not only can solve it, but also can obtained the sub-domain that values of it satisfy the given constraint.

## VI. CUNCLUDING REMARKS

This paper presents a new approach on path-oriented random testing that combines both the advantages of path testing and ART. We proposed a simple divide and conquer algorithm that permits to find efficiently sub-domain of a program variables domain that exercising a selected path. We obtain such sub-domain with partitioning. By applying a partitioning scheme on the input domain, the valid regions for test case generation can be easily identified. If such a favorable region cannot be identified, then the current partitioning scheme will be discarded and a refined one will be applied again.